# Leyenda: An Adaptive, Hybrid Sorting Algorithm for Large Scale Data with Limited Memory

Yuanjing Shi

*University of Illinois at Urbana-Champaign*

Zhaoxing Li

*University of Illinois at Urbana-Champaign*

## Abstract

Sorting is the one of the fundamental tasks of modern data management systems. With Disk I/O being the most-accused performance bottleneck [14] and more computation-intensive workloads, it has come to our attention that in heterogeneous environment, performance bottleneck may vary among different infrastructure. As a result, sort kernels need to be adaptive to changing hardware conditions. In this paper, we propose Leyenda, a hybrid, parallel and efficient Radix Most-Significant-Bit (MSB) MergeSort algorithm, with utilization of local thread-level CPU cache and efficient disk/memory I/O. Leyenda is capable of performing either internal or external sort efficiently, based on different I/O and processing conditions. We benchmarked Leyenda with three different workloads from Sort Benchmark [4], targeting three unique use cases, including internal, partially in-memory and external sort, and we found Leyenda to outperform GNU's parallel in-memory quick/merge sort implementations by up to three times. Leyenda is also ranked the second best external sort algorithm on ACM 2019 SIGMOD programming contest and forth overall. [1].

## 1   Introduction

Sigmod Programming Contest 2019 [1] is an annual programming contest host in parallel with the ACM SIGMOD 2019 conference. Student teams from degree-granting institutions are invited to compete to develop a fast sorting algorithm, which would be judged on the overall performance on a supplied workload. The key constraint here is that memory usage is limited, as only 30 GB of memory is allowed while the datasets vary in size from 10 GB to 60GB.

## 2   Background and Related Work

### 2.1   Sorting in Data-Intensive Systems

We take Apache Spark as an example to show what role sorting algorithm is playing in modern data management systems. Joins are the most computationally expensive among all SparkSQL operators [6]. With its distributed nature, Spark chooses proper strategies for join based on joining keys and size of data. Overall, there are three different join implementations, Broadcast Hash Join, Shuffle Hash Join and Sort Merge Join. Broadcast Hash Join is the first choice if one side of the joining table is broadcastable, which means its size is smaller than a preset ten-megabyte threshold. Although it is the most performant, it is only applicable to a small set of real world use cases. Between Shuffle Hash Join and Sort Merge Join, since Spark 2.3 [2], Sort Merge Join is always preferred over Shuffle Hash Join.

The implementation of Sparks' Sort Merge Join is similar to any other databases besides it happens on distributed partitions of data. In practice, the most expensive part of performing such a sort merge join operation is to arrange both the relations in sorted order and it is always assumed that the sizes of relations to be joined are large compared to the main memory and number of processors [11]. Hence, an explicit external sort operation is needed. In Spark, external sort is implemented in an AlphaSort manner [12], in which pairs of key-prefix and pointer are sorted instead of the actual records, which reasonably reduces data movement and random memory access. Only when the prefixes are equal will the actual records be retrieved via pointers and sorted based on comparison. During the sort process, a general implementation of Least-Significant-Digit Radix Sort may be used if it is supported by relations. After that, all records, in memory or spilled on disk, will be merged back.

### 2.2   Radix Sort and its Variations

Radix sort is a round-based non-comparative sorting algorithm that iterates through sorting keys in digits to separate them into buckets, grouping records by the individual digits, which share the same position and value. For example, records *"abc"*, *"bcd"*, *"cde"*, *"cfe"*, and *"dfg"* can be separated into 4 buckets according to the first character *"a"*, *"b"*, *"c"*, and *"d"* in the first round of radix sort. In the upcoming rounds, buckets will be further divided until all records are in sorted order. Each round of radix sort is of `O(n)` time complexity, and there can be at most `w` rounds, with `w` being the word length of the sorting key, so the overall time complexity of radix sort algorithm is `O(wn)`. Radix sort performs better than comparative sorting algorithms, like QuickSort or MergeSort, on workloads with fixed-size keys on each record [10].

| stable_sort | sort | **radix_sort** |
|---|---|---|
| 17.917s | 7.443s | **2.238s** |

Table 1: Preliminary Comparison on Sort Performance

Several comparison-based and non comparison-based sorting implementations on mutli-core processors have been proposed in the literature [8, 9, 13]. We first conduct some preliminary experiments on our first-version implementation of in-memory radix sort and GNU's implementations of parallel `sort` and `stable_sort`. Table 1 shows that Radix sort, as a non-comparison based sort that has a lower algorithmic complexity of O(N), remarkably outperformed GNU's sorting implementation on internally sorting 10 Gigabyte randomly uniform data.

However, radix sort has an irregular memory access pattern that results in misses to the cache and page hierarchy on modern CPUs, leading to wastage of memory band-width and increased memory latency. RADULS [9] is proposed to be able to handle skewed data by balancing loads between radix partitions/buckets but still leaves a huge memory fingerprint. And PARADIS, proposed by [8], addressed the memory issue with in-place parallel Least-Significant-Digit radix sort. However, one may find it difficult to fit these variations of radix sort into an external sorting pattern like the one required by Spark, as it imposes more I/O and memory constraints.

## 3   Leyenda

We address above issues with a novel external sorting algorithm - Leyenda, which is able to adaptively sort divergent workloads that may or may not fit into memory while leveraging an adaptive, parallel Most-Significant-Digit (MSD) Radix Merge-Sort implementation. Leyenda takes advantage of various I/O techniques, including direct I/O, increased parallelism with memory map and overlapping of computation and disk I/O by partitioning and balancing workloads in both sort and merge stages. Last but not the least, Leyenda also performs in-memory garbage collection and spawns threads with NUMA awareness to maximum memory bandwidth.

### 3.1   Adaptive Parallel MSD Radix Sort

As is mentioned above, the **O(wn)** time complexity makes radix sort a natural choice for sorting large-scale fixed-key-size sort data. Radix sort comes with two variations - Most-Significant-Digit (MSD), which starts sorting from the most significant digit of keys, and Least-Significant-Digit (LSD), which starts sorting from the least significant dight. Compared to LSD, MSD does not require the scattering process to continue until the last digit, thus can benefit from a speed up using comparison based sort for small buckets at the end. However, MSD radix sort requires each radix bucket to be sorted individually, which can be inefficient if the data is skewed, as small buckets will finish earlier than big buckets if they are sorted in parallel. Based on above analysis, and inspired by previous work [9, 13] on adaptive and parallel radix sort, we have proposed an **Adaptive Parallel MSD Radix Sort** algorithm which will load balance small buckets among threads, as well as breaking up big buckets to be processed by multiple threads:

---

**Result:** Sorted key[n]
**Data:** Unsorted key[n]
radix = 0; smallBucketQueue[];
buckets[256][k] = ParallelRadixSort(key, radix, n);
**if** *radix < 10* **then**
    **for** *bucket[k] in buckets[256][k]* **do**
        **if** *k > bigBucketThreshold* **then**
        │  MSDRadixSort(key, radix + 1, k);
        **else**
            smallBucketQueue.enqueue(task(key, radix + 1, k));
        **end**
    **end**
**end**
threads[40];
**for** *thread in threads[40]* **do**
    **while** *smallBucketQueue.dequeue(task)* **do**
        key, radix, n = task;
        buckets[256][k] =
          SingleThreadRadixSort(key, radix, n);
        **if** *radix < 10* **then**
            **if** *k < tinyBucketThreshold* **then**
            │  ComparisonSort(key);
            **else**
               smallBucketQueue.enqueue(task(key, radix + 1, k));
            **end**
        **end**
    **end**
**end**

**Algorithm 1:** MSDRadixSort

In the above algorithm, we create two thresholds `bigBucketThreshold` and `tinyBucketThreshold`. `bigBucketThreshold` is the threshold where the size of the bucket is large enough such that the benefit of parallelism out-weigh the overhead of threads, and `tinyBucketThreshold` is the threshold where size is small enough for comparison-based sort to perform more efficient than Radix Sort. This adaptive approach allows the algorithm to choose the best sort algorithm according to the key array size.

**Cache Aware Scattering Process**: For each round of MSD radix sort, we first compute a histogram based on the current `radix` to decide the size of each bucket and where each key should be placed. We then scatter the keys to their bucket offsets by copying from the `key` to the `tmp` array. Since this scattering process takes the most time in radix sort, it is beneficial for us to aggregate the keys in the thread level CPU cache first, before scattering them into the tmp array located in the main memory, which is similar to the LSD Radix Sort routine proposed in [13]. The MSD radix sort routine will then continue in each bucket recursively until it reaches the last `radix`, or when the bucket is small enough to run a comparative sort.

## 3.2 Effective I/O for Effective External Sort

As is shown in table 1, I/O takes a significant portion of the overall time, especially for in-memory internal radix sort, in which sorting takes 2 seconds and disk I/O takes 27 seconds. Thus it is necessary to optimize the I/O process based on the characteristics of the storage and underlying file system of OS. Leyenda has brought major optimizations on both internal and external workloads with a few novel and efficient I/O patterns, which correlate well with the MSD radix sort algorithm.

**Fast Direct Disk I/O**: Contrary to common myths which state that sequential read and write operations are strictly superior to random read and write operations, previous work shows in [7] that parallelism can significantly benefit both random read and write operations, and small random reads can achieve even better performance than single-threaded sequential reads if the queue depth (the number of concurrent jobs) is larger than 10. As a result, Leyenda reads data from disk in parallel with request size of 256KB, bypassing the OS page cache. This allows Leyenda to reach I/O speed faster than sequential read, as well as performing computation on data without interfering with read performance. Given this advantage, Leyenda extracts the `(key, pointer)` tuples from records during the read process. This allows radix sort to be performed on key-pointer tuples only, which is 10 times faster than sorting on the original record.

**Parallelism, Page Cache and Memory Map**: Since Direct I/O access to the disk will never exceed the bandwidth of the physical storage devices, it is inefficient to spill sorted data to disk if it is expected to be consumed by another process as fast as possible. Thus, if extra memory is available, Leyenda will always write sorted data to the OS page cache instead of persistent storage. Memory map (`mmap()`) is an API which provides applications a way to map a kernel address space to a user address space. With `mmap`, Leyenda maps the underlying disk file into page cache and perform concurrent I/O opera-tions on it. Combining this I/O pattern with the MSD radix sort, Leyenda is able to scatter data records directly to the page cache using `mmap()` in parallel, given the sorted `(key, pointer)` tuples. This reduces the disk write time by at most 50% and allows Leyenda to achieve state-of-art performance on in-memory workloads.

One flaw of writing aggressively to the page cache is that the OS will be taking care of writing data back to persistent storage while applications return with a successful flag on I/O operations. Theoretically, this process make I/O operations less durable as there may be loss of data if the OS level write-back is interrupted. There is an ongoing discussion within the Linux kernel community on how to handle write-back errors of page cache, but this flaw shall grow less notable as the disk bandwidth is catching up fast with memory bandwidth.

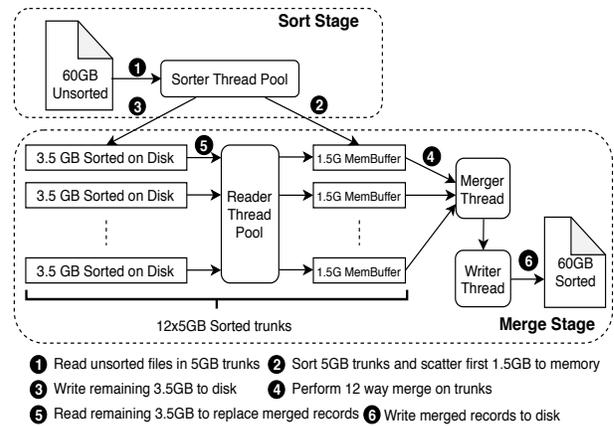

Figure 1: External Sort-Merge-Write Routine

**Overlapping Computation with I/O**: Memory map is the most useful while handling in-memory data or partially in-memory data. However, for external workloads, it is time for us to re-think the sort-merge pattern and how to leverage parallelism in such a pattern. We tackle the most expensive part of an external sorting with our radix sort implementation but merging those sorted partitions still takes a considerable chunk of the overall execution time.

To fully utilize memory and overlap I/O with computation, Leyenda parallelizes sort and merge stages by creating and coordinating between I/O threads and sort (merge) threads. As shown in figure 1, in the sort stage, Leyenda reads and sorts records in small trunks and writes them down to disk at the same time, creating several sorted intermediate files. In the merge stage, the sorted intermediate files need to be merged together as the final output. Leyenda will allocate a reader thread pool to continuously read data from disk, one partition at a time. Once a partition is read, it will be sorted and merged by the merger thread to a globally sorted partition, using a K-way merge algorithm. Then the writer thread will write it to the final output file. The partitioning, in other words - load balancing process, makes it possible for read/merge/write threads to work simultaneously and makes our external sort algorithm achieve the state-of-art performance on external workloads.

**Garbage Collection and NUMA Awareness**: One performance bottleneck we have observed while writing to the page cache is that applications cannot tell the operating system to free user space memory and use it as page cache for its own workloads. As a result, when the OS runs out of memory to spawn more pages, swapping will start which significantly hurts memory I/O bandwidth. To solve this problem, we have implemented a counter based garbage collector in C++ to aggressively free memory when it is no longer needed. Combined with a fine-grained memory allocation of 4KB buffers, Leyenda allows the OS to reclaim user space memory to create enough pages, supporting fast writing to the page cache. Another bottleneck is that the I/O threads, when allocated on a CPU which is not directly attached to the memory on a NUMA system, will lag behind other threads, significant delaying memory writing. Thus Leyenda will create I/O threads on the CPU local to the memory only to maximize memory bandwidth.

## 4 Evaluation

### 4.1 Experiment Setup

The workload consists of three different data sets requiring in-memory (10 GB), partial in-memory (20 GB) and external (60 GB) sort each. Each data set consists of a few hundreds of millions of 100-byte records and is generated by `sortbenmark.org`'s record generation tool [3]. Each record is made up of a 10-byte key and 90-byte value. Key distribution is set to uniform for 10 & 20 GB and skewed for 60 GB. Input data is binary for 10 & 60 GB datasets and ASCII for 20 GB dataset.

| Processor | 2x Intel Xeon E5-2640 v4 CPU |
|---|---|
| Configuration | 20 Cores / 40 Hyperthreads |
| Main Memory | 30 GB of main memory |
| Operating System | Ubuntu 17.10 |
| Storage | SATA III Solid State Disk |
| Software | CMake 3.9.1, gcc/g++ 7.2.0 etc. |

Table 2: Testbed Details

The final evaluation will take place on he testbed of the contest. Every submission is benchmarked in an isolated container and will be killed imminently if the memory limit is reached. The configuration is shown in Table 2.

### 4.2 Baseline Performance - GNU Sort

| Total | Read | Sort | Write |
|---|---|---|---|
| 60.280s | 20.743s | 17.917s | 14.273s |

Table 3: Baseline Performance - GNU::stable_sort

The start package provided by the contest is not efficient enough to conduct our baseline experiment as its execution time already exceeds the time limit of the first small dataset (10GB), which is around 300 seconds. We first build our baseline algorithm based on GNU's parallel sort algorithm [5] with a simple multi-thread Read/Write IO using *iostream*. Table 3 shows the performance of this baseline sorting on internal workloads (10 GB).

### 4.3 Overall Performance - Leyenda

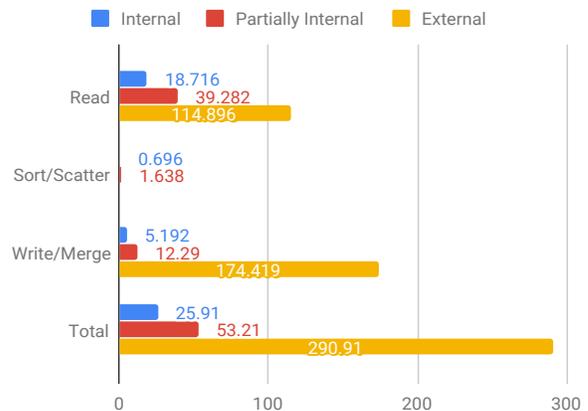

Figure 2: Overall Performance in seconds  Leyenda

1. **Sort Performance**: Excluding the time of scattering records to the page cache, Leyenda is able to sort 100 million key-pointer pairs in under one second and 200 million key-pointer pairs in around one and a half seconds. It is obvious that Leyenda make sorting no longer a dominant factor of the performance of an internal or external sorting algorithm. I/O becomes the bottleneck for external sorting in heterogeneous environment with various bandwidth characteristics on main memory and storage devices.

2. **I/O Performance**: Read performance is nearly linear on Leyenda, and is bound by the disk's maximum sequential read speed. As for write performance, since memory map and page cache help consequentially reduce write time on internal workloads, Leyenda performs way better than direct I/O access. Also, since sorting is overlapped with read and merging is overlapped with read/write on external sort, only read and merge time are reported for external workloads.

3. **Sigmod Programming Contest**: Leyenda is placed *fourth* on the overall ranking of the 2019 Sigmod Programming Contest with the second best performance on external workloads among more than 5000 submissions and 15 teams from all over the world which have submissions passed on external workloads. Notably, with minor revisions on order of file I/O requests and CPU affinity, Leyenda is able to outperform the state-of-art sorting performance on internal and partially internal workloads.

## 5 Future Work

1. **Integration with Data Intensive Systems**: Now that the sorting part of our algorithm is able to sort one million keys in under one second, not counting the memory scattering time, it is a good chance that Leyenda can be integrated with modern data intensive systems by complementing their sorting kernel as well as components that requires an efficient sorting algorithm. As mentioned above, modern distributed data processing/management systems like Apache Spark utilize sorting to advance and complete distributed tasks. As an adaptive and efficient sorting kernel, Leyenda can bring considerate performance improvement along with its flexibility on disk I/O access.

2. **Leyenda in the next step**: Right now Leyenda is able to sort records with fixed-size keys as a nature of its key sorting algorithm. To make Leyenda work in more general use cases, the next step of Leyenda's development will be making it work on flexible key sizes. Either can we construct a prefix generator for all types of keys or make Leyenda hybrid between prefix sorting and comparison-based record sorting, like the sorting pattern used in Spark's Sort Merge Join.

## Acknowledgments

We'd like to thank Illinois Engineering IT for providing the test VMs.

## Availability

Our work shall be available as a Github repository at: github.com/shingjan/leyenda by the time ACM SIGMOD 2019 programming contest finalizes its leaderboard.